\documentclass[fleqn,twoside]{article}
\usepackage{espcrc2}

\title{CONFINEMENT-DECONFINEMENT ORDER PARAMETER
 AND DIRAC'S QUANTIZATION CONDITION}

\author{P.A. Marchetti\address{Dipartimento di 
Fisica, Universit\`a di Padova and INFN-Sezione di Padova, I-35131 Padova, 
Italy}\thanks{This work was supported by the European Commission RTN 
programme HPRN-CT2000-00131.}}

\begin{document}

\begin{abstract}

We describe a monopole-like order parameter for the confinement-deconfinement transition 
in gauge theories where dynamical charges and monopoles coexist. It has been recently proposed in a collaboration with J. Fr\"ohlich.
It avoids an inconsistency in the treatment of small scales present in earlier definitions of monopole fields by respecting Dirac's quantization condition 
for electromagnetic fluxes.
An application to $SU(2)$ lattice Yang-Mills theory is outlined, naturally fitting in the 't Hooft scenario for confinement.
\vspace{1pc}
\end{abstract}

\maketitle
\section{Monopoles and confinement}

't Hooft proposed to explain confinement in $SU(2)$ Yang-Mills theory as a consequence of condensation of magnetic monopoles defined as follows. He suggested \cite{'th} to 
construct a scalar field $X(U)$ with values in $su(2)$, as a function 
of the gauge field $U$ and transforming in the adjoint representation of the 
gauge  group $SU(2)$. By requiring that $X(U)$ be diagonal one then fixes a gauge (``Abelian projection"). The resulting theory exhibits  a residual $U(1)$ 
gauge invariance.

The argument of the diagonal component of the $SU(2)$ gauge field in this 
``Abelian projection gauge" plays the role of a compact $U(1)$ ``photon" field, 
$A_\mu$, with range ($-2\pi, 2\pi$), and the off-diagonal components are 
described by a complex field, $c$, charged with respect to the residual 
$U(1)$ gauge group. The points in space-time where the two eigenvalues of the 
matrix $X$ coincide identify the positions of the monopoles in this gauge. 
Confinement is believed to emerge as a consequence of 
monopole--condensation in the form of a ``dual Meissner effect''.

This characterization of  confinement needs a charged order parameter in presence of both dynamical charges and monopoles, in the dual Higgs model. 

\maketitle
\section{Monopoles in Abelian gauge theories}

\quad
In scalar electrodynamics, or non-compact Higgs model, with charged scalar field $\phi$ one can construct 
gauge--invariant charged fields 
adapting the Dirac recipe \cite{di}, dressing the local non--gauge invariant field 
$\phi (x)$ with a cloud of soft photons described by a
phase factor with argument given by the photon field $\vec A$ weighted by a
classical Coulomb field $\vec E^{\vec x}$ generated by a charge at ${\vec x}$. More precisely, we define an electric distribution $
E_\mu^x(y)=\{0, {\vec E}_i^{\vec x} (\vec y) \delta(x^0-y^0)\}$ satisfying $\partial^\mu E_\mu^x (y)= \delta (x-y)$ and the non-local gauge-invariant charged field is given by $
\phi(x) e^{i \int  A^\mu E_\mu^x}$.
One can prove rigorously in the lattice approximation that the vacuum expectation value $
\langle \phi(x) e^{i \int  A^\mu E_\mu^x} \rangle \neq 0 $ in the Higgs phase and it vanishes in the Coulomb phase \cite{fm}.

In Abelian gauge theories there is a natural notion of duality exchanging 
the r\^ole of charges and monopoles \cite{dual}. In particular, one can obtain monopole correlation 
functions from gauge--invariant charged correlation functions by a duality 
transformation. The non-compact Higgs model $(n.c.H.)$ with charges and photons (described by $A_\mu$) is mapped by duality to the $U(1)$ gauge theory with monopoles and dual photons (described by $\tilde A_\mu$) and the charged 
correlator  
\vfill\eject
\begin{equation}
\label{char}
\langle \bar\phi(x) e^{-i \int  A^\mu E_\mu^x} \phi(x') e^{i \int  A^\mu E_\mu^{x'}} \rangle_{n.c.H.} 
\end{equation}
is mapped into the disorder field correlator 
\begin{equation}
\label{didu}
\langle e^{-[ S_{U(1)} (\partial_{[\mu} {\tilde A_{\nu]}} + \partial^\rho \Delta^{-1} B^{xx'}_{\mu\nu\rho})-S_{U(1)}(\partial_{[\mu} \tilde A_{\nu]})]} \rangle_{U(1)}
\end{equation}
where $B^{xx'}_{\mu\nu\rho}= B^{xx}-B^x+\omega^{xx'}$ with
$\Delta$ denoting the Laplacian, $B^x_{\mu\nu\rho}=\epsilon_{\mu\nu\rho\sigma}E^x_\sigma,$ and $ \omega^{xx'}_{\mu\nu\rho} $ is an integer 3-current from $x$ to $x'$ required by magnetic  flux conservation in $U(1)$ gauge theory and in this respect playing the role of the role of $\langle \bar\phi(x') \phi(x)\rangle_{n.c.H.} $ in eq.(\ref{char}).

In the models previously considered there were only dynamical charges or monopoles. Let us consider the changes needed in models where dynamical charges and monopoles coexist, like the compact Higgs model.
 In this model, the Dirac surfaces, $S$, swept by the Dirac strings of monopoles  are described by integer-valued  surface currents, $n^{\mu\nu}$ Hodge dual to $S$. A change of Dirac surfaces, $S\rightarrow S^\prime$, for a fixed configuration of monopole worldlines, corresponds to the shift 

\begin{equation}
\label{shi}
n^{\mu\rho} \rightarrow n^{\mu\rho}+ \partial^\mu V^\rho - \partial^\rho V^\mu, 
\end{equation}
where $V^\mu$ is the integer current Hodge dual to the volume whose boundary is the closed surface $S^\prime - S$.
In the partition function, the interaction of the electric currents generated by the charged particles, $j_\mu$, with the Dirac surfaces of the monopoles is of the form

\begin{equation}
\label{int}
ieg \int j_\mu \partial_\rho \Delta^{-1} n^{\rho\mu} 
\end{equation}
where $e$ is the electric charge of the matter field and $g$ the magnetic charge of the monopole field.
The change (\ref{shi}) induces a shift of (\ref{int}) by

\begin{equation}
\label{dico}
i e g \int j_\mu V^\mu
\end{equation}
which when exponentiated is unity, as physically required, provided (\ref{dico}) is an integer multiple of $2\pi i$ [Dirac quantization condition for fluxes]. This happens in the partition function if Dirac's quantization condition for charges holds, i.e.  $e g = 2\pi q, q$ an integer, because $j_\mu$ and $V^\mu$ are integer currents. In the Dirac ansatz for the 2-point function of the charged field, however, $j_\mu$ acquires additional Coulomb-like terms, $E_\mu$, which are real-valued. The action then acquires a monopole-charged field interaction term 

\begin{equation}
eg \int E_\mu V^\mu  \notin 2\pi {\bf Z}  
\end{equation}
even if $e g \in 2\pi {\bf Z}$, and the Dirac strings of monopoles become unphysically ``visible".
An obvious cure for this inconsistency  would be to replace the Coulomb field $E^x_\mu$ by a ``Mandelstam string" $j_\mu^x$ \cite{man},  squeezing the entire flux of $E^x$ into a single line from $x$ to $\infty$ at fixed  time (and adding suitable b.c.).

However, this squeezing of the flux is so strong that it produces IR divergences [with a lattice UV cutoff $\int (E^x_\mu - E_\mu^{x'}) \Delta^{-1} (E^{x \mu} - E^{{x'} \mu}) < \infty$ but $\int (j^x_\mu - j^{x'}_\mu)\Delta^{-1}(j^{x \mu} - j^{{x'} \mu}) = \infty $].

To avoid these divergences, we propose \cite{fm1} to replace a fixed Mandelstam string by a sum over fluctuating Mandelstam strings $j_\mu^x$ weighted by a measure ${\cal D} \nu_q (j_\mu^x)$ with the property that in the scaling limit,
\begin{equation}
\label{cure}
\nonumber
\int {\cal D}\nu_q (j^x_\mu) e^{i e\int j_\mu^x A^\mu}\sim e^{i e \int
E_\mu^x A^\mu}.
\end{equation}
[The integer $q$ in the measure ${\cal D} \nu_q$ is the one appearing in the Dirac quantization condition $eg= 2\pi q$].
 A measure with such property is the measure over ${\bf Z}/q$-valued currents appearing in the Fourier representation in terms of the gauge field $A_\mu$ of the spin correlator of a 3D gauged Villain model with period $2 \pi q$, in the broken symmetry phase, with a point removed at infinity. This measure is supported on currents $ j^x_\mu$ associated with $q$ 
paths in a 3-plane at a fixed time, starting at the site $x$ and reaching a common point at infinity . From (\ref{cure}) we see that the 
measure  ${\cal D} \nu_q (j^x_\mu)$ is peaked at $E^x_\mu$ at large scales. 

The 2-point correlation function for the gauge-invariant charged field in the compact Abelian Higgs ($c.H.$) model has then the form 
$$
\int {\cal D}\nu_q (j^x_\mu) \int {\cal D}\nu_q (j^{x'}_\mu) \langle 
\phi (x) \bar\phi (x')
e^{i e \int (j^x_\mu - 
j_\mu^{x'}) A^\mu } \rangle_{c.H.}
$$
\begin{equation}
\label{ch}
\\
\end{equation}
replacing the correlator (\ref{char}) of the non-compact model. This definition respects Dirac's quantization condition for fluxes and, as a consequence, it is independent of the Dirac strings of the magnetic monopoles of the compact Higgs model. In \cite{chord} one finds a numerical evidence for the validity of an order parameter for the Coulomb-Higgs transition in this model, based on the above correlation function (using the formalism of effective potential).

The 2-point monopole correlation function obtained by duality from (\ref{ch}) is given by  
\begin{equation}
\label{mord1}
\int {\cal D} \nu_q (j^x_\mu) \int {\cal D}\nu_q (j^{x'}_\mu)
\langle D(\Sigma (j^x-j^{x'}+j^{xx'})) \rangle{\tilde {\phantom a}}
\end{equation}
where $j^{xx'}$ is the dual of $\omega^{xx'}$ (see eq.(\ref{didu})). Here $D(\Sigma)$ is the 't Hooft loop \cite{dual} in the dual of the compact Higgs model $({\tilde {\phantom a}})$. The surface $\Sigma$ has boundary given by the support of $ j^x-j^{x'}+j^{xx'}$, with b.c. turning it into a closed curve.
$D(\Sigma)$ is obtained by shifting the field strength of $A_\mu$ by $2 \pi q * \Sigma_{\mu\nu}$ in the action, where $q \Sigma_{\mu\nu}$ is 
the ${\bf Z}$-valued surface current supported on $\Sigma$ and * the Hodge dual. Since $j^x_\mu $ 
is supported on $q$ paths, $\Sigma$ is a $q$-sheet surface with the $q$ sheets having a common boundary given by the single line support of $j^{xx'}_\mu $.
\maketitle
\section{Monopoles in Yang-Mills theory}

We wish to export the above ideas to $SU(2)$ lattice Yang-Mills theory, with action defined by 

\begin{equation}
\label{ym}
S_{YM}(U_{\mu\nu}) = - \beta \sum_{y,\mu,\nu}
Tr U_{\mu\nu} (y).
\end{equation}
where $ U_{\mu\nu} (y)$ is the Wilson plaquette with initial point the site $y$.
Firstly one remarks \cite{fm2} that, in an Abelian projection gauge, there appear a charged field, $c$, of 
electric charge 1 and regular \cite{reg} monopoles with magnetic charge $g=4 \pi$, whose condensation should be responsible for confinement, and for them Dirac's quantization condition for charges is satisfied with $q=2$. 

Integrating out $c$ in the partition function one obtains an effective action $U(1)$- gauge invariant $S^X_{U(1)}(\partial_{[\mu} A_{\nu]})$. Since $A$ is 4 $\pi$ periodic, a Fourier expansion yields:
\begin{equation}
e^{-S^X_{U(1)}(\partial_{[\mu}A_{\nu]})}= \int {\cal D} \ell^{\mu\nu} e^{{i \over 2} \int \ell^{\mu\nu} \partial_\mu A_\nu} F(\ell^{\mu\nu})
\end{equation}
where $\ell^{\mu\nu}$ is an integer surface current, for a suitable functional $F$.
 Integrating out $A_\mu$ one obtains  $\partial_\nu \ell^{\nu\mu} =0$.
We replace the integer currents $\ell^{\mu\nu}$ by a real gauge field strength $\epsilon^{\mu\nu\rho\sigma} \partial_\rho\tilde A_\sigma$ solving the above constraint and use a  Fourier representation of the integrality condition (Poisson formula) in terms of integer currents $\rho^\mu$, describing the Abelian projection monopole worldlines. This yields a representation of the partition function of the dual theory:

$$
\tilde Z =\int {\cal D} \tilde A_\mu F(\epsilon^{\mu\nu\rho\sigma} \partial_\rho \tilde A_\sigma) \int {\cal D} \rho^\mu  e^{i 4 \pi \int \rho^\mu  \tilde A_\mu},
$$
with $\partial_\mu \rho^\mu =0$, by $U(1)$ gauge invariance.
The 2-point correlator for the gauge-invariant charged field in the dual model is given by:
$$
\int {\cal D} \nu_2 (j^x_\mu) \int {\cal D} \nu_2 (j^{x'}_\mu) \langle e^{i 4 
\pi \int (j^x -j^{x'} + j^{xx'})_\mu \tilde A^\mu} \rangle{\tilde {\phantom a}}
$$
where
$j^x$ is a 2-path half-integer current at constant time  and $j^{xx'}$ an integer current required by flux conservation as $\omega^{xx'}$ in eq.(\ref{didu}).
Applying backward the duality transformation one obtains eq.(\ref{mord1}) for $q=2$ and action $S^X_{U(1)}$.
 Reexpressing this correlator in terms of the original $SU(2)$ gauge field $U$ yields:
$$
\int {\cal D} \nu_2 (j^x_\mu) \int {\cal D} \nu_2 (j^{x'}_\mu) \langle D(\Sigma (j^x -j^{x'} + j^{xx'})\rangle_{YM}
$$
\begin{equation} 
\label{S}
=\langle {\cal M}(x) {\cal M}(x') \rangle_{YM}.
\end{equation}

Here $D(\Sigma)$ is the 't Hooft loop in $SU(2)$, obtained substituting in the Yang-Mills action (\ref{ym}) $U_{\mu\nu}$by $ U_{\mu\nu} e^{i 2 \pi \sigma_3 \Sigma_{\mu\nu}}$ and
$x,x'$ are the creation and annihilation points for the monopole field ${\cal M}$.
The definition in (\ref{S}) is independent of the choice of an Abelian projection, whereas the  initial definition of monopole currents $\rho$ was projection-dependent. Hence the position where the monopoles are created or annihilated are intrinsic to the $SU(2)$ Yang-Mills theory, but to define the trajectories of the monopoles one needs an Abelian projection; these monopole do not appear to have a semiclassical limit.  
We propose \cite{fm2} the following criterion for confinement based on monopole condensation:
\begin{equation}
\langle {\cal M}(x) {\cal M}(x') \rangle_{YM} \rightarrow_{|x-x'| \rightarrow \infty} c > 0.
\end{equation} 

A justification for our criterion is based upon the following considerations: 

1) since $2 \Sigma_{\mu\nu}$ is integer valued in the 't Hooft loop we can substitute 
\begin{equation}
e^{i 2 \pi \sigma_3 \Sigma_{\mu\nu}} \rightarrow e^{i 2 \pi X \Sigma_{\mu\nu}}
\end{equation} 
for any choice of $X$ selecting an Abelian projection\\
2) since the measure $ {\cal D} \nu_2 (j^x_\mu)$ is peaked at large scales around  $E^x_\mu$, in a mean-field approximation with respect to
$\int {\cal D} \nu_2 (j^x_\mu) \int {\cal D} \nu_2 (j^{x'}_\mu)$ we have in the scaling limit 
$$
\langle \Sigma_{\mu\nu}\rangle_{j^x j^{x'}} \simeq  \partial^\rho \Delta^{-1} B^{xx'}_{\mu\nu\rho}.
$$

Hence in the above mean field we have: 

\begin{eqnarray}
\label{x}
&\langle {\cal M}(x) {\cal M}(x') \rangle_{YM} \simeq \nonumber\\ 
&\langle e^{-[S_{YM} (U_{\mu\nu} e^{i X 2 \pi 
\partial^\rho \Delta^{-1} B^{xx'}_{\mu\nu\rho}}
)-S_{YM}(U_{\mu\nu})]}\rangle_{YM} \nonumber \\
&=\langle {\cal M}_{MF}(x) {\cal M}_{MF}(x') \rangle_{YM}.   
\end{eqnarray}

This is the order parameter proposed by Di Giacomo et al. \cite{mord} (see also \cite{pol} for a variant). The ``Mean-Field'' v.e.v. $\langle {\cal M}_{MF} \rangle_{YM}$ is numerically a good order parameter for the confinement-deconfinement transition \cite{dig1}. Nevertheless the definition (\ref{x}) is inconsistent in the treatment of small scales because it violates Dirac's quantization for fluxes and therefore it depends on the choice of Dirac strings [at order $\epsilon^2 , \epsilon$ lattice spacing, assuming good continuum limit for the fields of the Abelian projection]. However in our approach it is simply the  Mean-Field of a correlator well defined even at small scales, strictly independent of Dirac strings and Abelian projection.

{\bf Acknowledgments.} It is a pleasure to thank J. Fr\"ohlich for the joy of a long collaboration. Useful discussions with M. Chernodub, A. Di Giacomo and M. Polikarpov are gratefully acknowledged.

\end{document}